\documentclass{aa} % for a referee version
\usepackage{graphics}
\begin{document}
\def\Tef{T$_{\rm eff}$}

   \thesaurus{07(08.01.1; 08.16.4; 08.03.1)} 
   \title{The formation of lithium lines in the atmospheres of super Li-rich AGB stars}

   \subtitle{}

   \author{C. Abia\inst{1}, Y. Pavlenko\inst{2}, P. de Laverny\inst{3}}
 
  \offprints{C. Abia}

   \institute{Dpt. F\'\i sica Te\'orica y del Cosmos, Universidad de Granada, E-18071 Granada, Spain\\
	    \and
	     The Main Astronomical Observatory of NAS, Golosiiv woods, 252650 Kyiv-22, 
	     Ukraine\\
	     \and
	     Observatoire de la Cote d'Azur, Dpt. Fresnel UMR 6528, B.P. 4229, F-06304 Nice Cedex 04, France\\
	     email: cabia@goliat.ugr.es; yp@mao.kiev.ua; laverny@obs-nice.fr}

\date{Received; accepted}
\authorrunning{Abia et al.}
\titlerunning{Li formation in SLiR stars}

\maketitle

\begin{abstract}
The formation of lithium lines in the atmosphere of C-rich giants is discussed. 
LTE and NLTE approximations are used to model lithium lines in the spectra of super Li-rich 
AGB stars. The system of equations of the statistical balance of neutral Li in plane-parallel 
model atmospheres is solved for a 20-level atom model. JOLA and line-by-line models of molecular 
absorption are used to compute synthetic spectra as well as the opacity in the 
frequencies of bound-bound and bound-free transitions of the lithium lines. Curves of 
growth and synthetic spectra are computed in LTE and NLTE for several model atmospheres of different 
\Tef~and C/O ratios for four lithium absorptions, namely: the $\lambda$4603, $\lambda$6104, 
$\lambda$6708 and $\lambda$8126 {\AA} Li I lines. The sensitivity of NLTE effects on \Tef~and the
C/O ratio is discussed. We found that NLTE mainly affects the resonance line doublet 
($\lambda6708$) while the impact of NLTE effects on the lithium subordinate lines, formed in the inner 
regions of C-rich giant atmospheres, is rather weak. Therefore the use of these lines is recommended for Li 
determination in AGB stars. However, in no case can we properly account for the formation of Li lines in AGB stars 
until sphericity, velocity stratifications, dust, chromospheres and other related phenomena, which are in fact 
present in AGB star atmospheres, are considered. Our results are used to derive Li abundances in three super-Li rich 
C-stars taking into account NLTE effects. Finally, the net Li yield from this class of stars into the 
interstellar medium is reconsidered.

\end{abstract}

%
%________________________________________________________________
\section{Introduction}

 The element lithium has provided many clues concerning stellar evolution and nucleosynthesis, as
well as cosmology. However, its origin is still far from clear. Nowadays there is an
increasing belief that this element has a multi-source nature: primordial nucleosynthesis
(see the review by Wallerstein et al. 1997), galactic cosmic ray spallation in the
interstellar medium and/or in the neighbourhood of supernova remnants (Meneguzzi et al.
1971; Feltzing \& Gustafsson 1994), late type stars and novae (Abia et al. 1993a; Hernanz et
al. 1996), supernovae explosions (Woosley \& Weaver 1995) and even spallation reactions
around compact objects (Guessoum \& Kazanas 1998). However, except perhaps for the Li 
production during the Big Bang there is no consensus about the contribution of these
sources to the present cosmic abundance, log $\epsilon$(Li)$\approx 3.3$\footnote{The abundance of a given
element X is noted as log $\epsilon$(X)$\equiv 12 +$log(N(X)/N(H)) where N(X)/N(H) is the abundance 
by number of the element X.}

The discovery of an unusually strong Li line at $\lambda6708$ {\AA} by McKellar (1940) in the carbon
star (C-star) WZ Cas strongly suggested that AGB stars might be an important source of Li in the
galaxy. The work of McKellar was followed by similar discoveries in other AGB stars: the C-stars
WX Cyg and IY Hya (Sanford 1950; Abia et al. 1991), the S-stars T Ara and T Sgr 
(Feast 1974; Boesgaard 1970) and the SC-stars Henize 166 and VX Aql (Catchpole \& Feast 1971; Warner 
\& Dean 1970). The same figure has also been reported in AGB stars of the Magellanic Clouds
(Plez et al. 1993; Smith et al. 1995). The pioneering analysis by Cohen (1974) showed that the
measured equivalent width of the $\lambda6708$ {\AA} Li I line (W$_{\lambda}\sim 1-10$ {\AA}) in these
stars is mainly the consequence of a strong enhancement of the lithium abundance in the atmosphere. 
Quantitative determinations by spectral synthesis (Denn et al. 1991; Abia et al. 1993b; Plez et al.
1993; Smith et al. 1995) have shown that the abundance of Li in these stars is 1-2 orders
of magnitude higher than the present cosmic Li abundance. Therefore, these stars have been named 
super lithium-rich (SLiR) stars and might well constitute the main source of Li in the galaxy. 
The production of Li in AGB stars has received special attention in recent years
(e.g. Sackmann \& Boothroyd 1992). Basically, Li is produced by the reaction
$\rm{^3He(^4He,\gamma)^7Be}$, followed by $\rm{^7Be(e^-,\nu)^7Li}$ in a hot convective
region that brings the $\rm{^7Be}$ or $\rm{^7Li}$ to cooler regions before the $^7$Li
is destroyed by $\rm{^7Li(p,\alpha)^4He}$. This is the so-called $^7$Be-transport
mechanism (Cameron \& Fowler 1971). Results of these theoretical studies quantitatively
agree with the Li abundances derived in AGB stars of the galaxy and the Magellanic Clouds,
although there are still many open questions related to the Li production in these stars. Among 
others we find questions about the minimum initial stellar mass which can eventually become 
a SLiR star during the AGB phase, the duration of the SLiR phase or the actual Li yield into the 
interstellar medium.

Our purpose in this work is related to the reality of the Li abundances in SLiR stars and its 
consequences on the Li yield. At present, the uncertainty in the derivation of the Li abundances in 
AGB stars is not lower than 0.4-0.5 dex (see references above), mainly due to uncertainties in the 
stellar parameters and the fit to the observed spectra. Note, that in many situations no single 
set of stellar parameters is found to fit the observations. However, there are also
systematic errors that are not usually taken into account that might dramatically change the Li abundance
derived: uncertainties in the atmosphere models, the existence of velocity stratifications
(most SLiR stars are actually variable), sphericity and NLTE effects are not currently considered
as possible systematic sources of error. Of course, the consequences of each of these phenomena
on Li abundance merit individual study and are beyond the scope of this
work (see, however, Scholtz 1992; J\o rgensen et al. 1992). Here, we will focus our attention
on the effects of departures from LTE in the formation of the Li lines in C-stars.

The study of NLTE effects in the formation of lithium lines in stellar atmospheres was begun by the work of 
M\"uller et al. (1975). They performed NLTE analysis of the very weak lithium resonance line $\lambda
6708$ {\AA} in the spectrum of the Sun. Later, Luck (1977) investigated the statistical balance of lithium 
in the atmospheres of G-K giants for a 4-level atom model of Li. A similar atom model was used by
de la Reza \& Querci (1978) and de la Reza et al. (1981). Historically, these papers considered for the first time 
the impact of a stellar chromosphere on the lithium lines. A new stage of research was begun with the work of Steenbock 
\& Holweger (1984), who used the technique of complete linearization for an 8-level atom model. 
They studied NLTE effects in lithium lines in atmospheres of dwarfs and giants. Pavlenko (1991) considered 
in detail the effects of deviation from LTE in the atmosphere of red giants. Later, Magazz\'u et al. (1992),
Martin et al. (1994) and Pavlenko (1994), continued the NLTE studies in T-Tau stars, G-K giants, subgiants 
and dwarfs. The main results of these studies were confirmed by the independent work of Carlsson et al. (1994), 
who made a similar investigation using atmosphere models with various effective temperatures, luminosities 
and metallicities. Finally, Houdebine et al. (1995), Pavlenko et al. (1995), Pavlenko \& Magazz\'u (1996) 
and Martin et al. (1997) have continued the studies of different aspects of NLTE formation of 
lithium lines in stellar atmospheres.  

As far as we know, the sole study on this subject for AGB stars is that by de la Reza \& Querci (1978) who
performed kinetic equilibrium calculations of neutral lithium lines in C-stars 
and determined the influence of the possible chromospheric radiation into the photosphere.
In the present work we revise this study using up to date atomic data, collisional and
radiative rates, an extended Li atom model and more reliable model atmospheres for C-stars in a wider 
range of effective temperatures (T$\rm{_{eff}}=2500-3100$ K) and C/O ratios (1.0-1.35) (see below). We explicitly 
apply our results deriving Li abundances from synthetic spectra, both in LTE and NLTE, in three well known 
SLiR stars (WX Cyg, WZ Cas and IY Hya) from four accessible Li I lines: the resonance line at $\lambda6708$ {\AA} 
and the subordinate transitions at $\lambda4603$, $\lambda6104$ and $\lambda8126$ {\AA}, respectively, benefiting from
the high signal-to-noise ratio and high resolution spectra of these stars.
The consequences on the real Li abundances in AGB stars and on their net Li yield into the 
interstellar medium is then reexamined.

\section{Observations}
The observations were made during 1997 and 1998 in two different observatories. We used the
4.2 m WHT at the Observatory of El Roque de los Muchachos with the Utrecht Echelle Spectrograph
as the main instrument and a 2048$\times$2048 CCD with 24 $\mu$m pixel size. We
used the 79.0 lines/mm grating which provides less wavelength coverage, but more space between orders
(20-30 arcsec). The projected size of the slit on the chip was around two pixels which gave
a resolving power of 50000, the effective resolution ranging between 0.05-0.19 {\AA} from the blue
orders to the red ones. The total number of orders on the chip were 30 covering the wavelength
range 0.4-1.0 $\mu$m with some gaps between orders. WZ Cas and WX Cyg were also observed by the 2.2
m telescope at the Calar Alto Observatory. For this observational run a fibre optics cassegrain
echelle spectrograph (FOCES) (Pfeiffer et al. 1998) was used. This time the chip was a 1024$\times$1024 
Tektronik CCD with 24 $\mu$m pixel size. The FOCES image covers the visible spectral region from 0.38 to
0.96 $\mu$m in about 80 orders with full spectral coverage. Spectral orders are separated by 20 pixels in the
blue and 10 in the red. The maximum resolving power is 40000 with a two pixel resolution element.
%__________________________________________________ Two column table
\begin{table*}
\caption[]{Log of the observations and stellar parameters}
{!}{\includegraphics{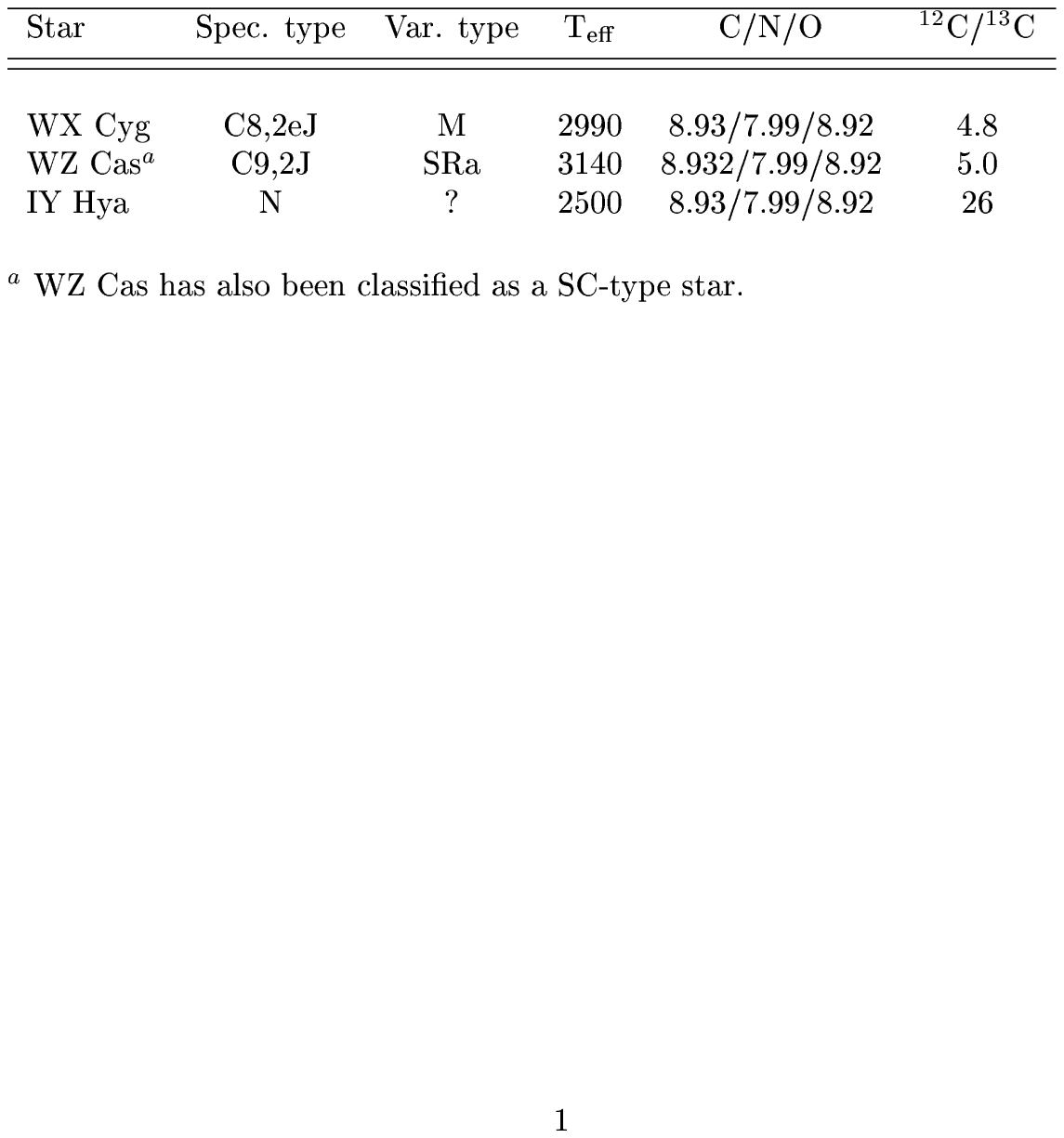}}
\end{table*}

%___________________________________________________________________

The reduction of the spectra was made following the standard procedures using the ECHELLE task of
the IRAF software package: bias subtraction, division by flat-field images, removal of the scattered-light, 
extraction of the orders and wavelength calibration with Th-Ar lamps. Typically the rms in the residuals
of the calibrations were better than 10 m\AA. Finally, the spectra were divided by the
spectrum of a hot rapidly rotating star to remove telluric absorptions, although they are only
important in the $\lambda8126$ {\AA} range. The signal-to-noise ratio of the spectra vary along the
wavelength, from S/N$\sim 20$ in the bluest orders to S/N$\sim 700$ around $\lambda8000$ {\AA}.
Since our stars are relatively bright objects (V$\sim 7-11$), this strong variation in the
S/N ratio achieved is mainly due to the very low emissivity of the stars below $\lambda\sim 4500$ \AA.
This is particularly evident in IY Hya, which is an extremely red object. For this reason, its spectrum
in the $\lambda4603$ {\AA} range is of too low quality (S/N$\sim 30$) and has not been considered.

\section{The abundance analysis}
\subsection{Atomic and molecular line lists}
The line list described in L\`ebre et al. (1999) completed with C$_2$ lines has been used for the analysis 
of the $\lambda6708$~\AA\ region. Lines from the $^{12}$C$_2$, $^{12}$C$^{13}$C and $^{13}$C$_2$ 
Swan system (Kurucz 1998) were considered and from the Phillips red system were predicted as in 
de Laverny \& Gustafsson (1998). The line lists 
for the $\lambda4603, \lambda6104$ and $\lambda8126$ {\AA} spectral domains were built in a similar 
way for the C$_2$ and CN molecules, and atomic line data were found in VALD (Piskunov et al. 1995). 
The broadening by radiation and van der Waals damping were calculated as in de Laverny \& 
Gustafsson (1998). Line data were adjusted as described in L\`ebre et al. (1998) by comparison 
with the solar spectrum using the Holweger \& M\"uller (1974) model atmosphere for the Sun with
element abundances from Anders \& Grevesse (1989). A good fit to the solar spectrum was obtained in the
$\lambda6708$ and $\lambda8126$ {\AA} spectral domains in $\sim 50$ {\AA} around the corresponding Li line.
The fit, however, was not as good in the other two spectral ranges. Indeed, in the $\lambda4603$ and
$\lambda6104$ {\AA} regions there are a number of unidentified features in the solar spectrum with a modest 
absorption ($\sim 5-25$ m{\AA}). Most of these features could be atomic in nature because no such intense
molecular absorptions are expected in the Sun in these spectral ranges. However, we did not
find any atomic lines in the data base of VALD or of Kurucz (1998) at the wavelengths of the
missing features. Nevertheless, we were able to obtain a good fit to the solar spectrum
in at least 10 {\AA} around the $\lambda4603$ {\AA} and $\lambda6104$ {\AA} Li I features. Thus, 
we do not believe this problem will introduce an additional source of error into the abundance of Li derived 
from these two spectral domains.

\subsection{Atmospheric parameters}
Stellar parameters for the stars studied were taken from the literature when available.
For WX Cyg the effective temperature derived by Ohnaka \& Tsuji (1996) from the infrared
flux method (IRFM) was adopted. For WZ Cas, effective temperature derived from the IRFM and
angular radii measurements (Dyck et al. 1996) agree quite well ($\pm 30$ K). However, the
value derived from these two methods (3150 K) contrasts with that derived from infrared 
photometry ($\sim 2800$ K; see Noguchi et al. 1981; Frogel et al. 1972). We adopted here 
the value obtained by Dyck et al. (1996). For IY Hya there is no estimate of its effective 
temperature nor its photometry is available in the literature. We estimated the effective temperature 
by comparing its spectrum with the temperature sequence spectral atlas for C-stars created by Barnbaum et al. 
(1996). From that comparison we believe that IY Hya is a cool N-star and consequently, we adopted the lower 
effective temperature in our grid of atmosphere models i.e.: T$_{\rm{eff}}=2500$ K. 
The uncertainty in T$_{\rm{eff}}$ is, however, no less than $\pm 200$ K; it might well be 
larger for IY Hya. Note also that these stars are variable and we suspect \Tef~variations as large
as $\sim 300$ K during their cycle (see Richichi et al. 1995). The consideration of an unique effective
temperature for these stars is therefore a strong assumption.

A solar metallicity ([Fe/H]=0.0) was considered for the three stars. Most galactic 
C-stars are of Population I with near solar metallicity although typically they show strong 
enhancements of heavy elements (Zr,Ba,La etc...)(Utsumi 1985; Dominy 1985). We checked that 
there is no intense heavy element line close enough to the four Li absorptions in our line lists 
that may affect the Li abundance derivation. A study of the heavy element enhancements in the SLiR stars 
and other normal C-stars will be presented in a separate work. 

A gravity of log g$=0.0$ and a microturbulence 
parameter of $\xi=2.5$ kms$^{-1}$ were adopted. These are typical values for C-stars (see Lambert et al. 
1986). Concerning gravity, note that studies by Pavlenko (1990) and Magazz\'u et al. (1992) on O-rich 
dwarfs, subgiants and giant stars showed no important sensitivity in the ratio of LTE/NLTE Li abundances 
caused by changes in the gravity of $\Delta$log g$=\pm 1$, especially in the case of strong lines.
Thus, uncertainties in gravity play a minor role in the formation of the lithium 
lines in C-rich atmospheres. Radiative damping (which does not depend on log g) plays the 
main role in the formation of the wings of the saturated lithium lines. More serious seem to be the 
consequences of changes in the temperature structure of the atmosphere with decreasing gravity due to 
the increase in the effectiveness of sphericity effects. However, analysis of these effects lies 
beyond the scope of this paper.

\begin{figure*}
\resizebox{\hsize}{!}{\includegraphics{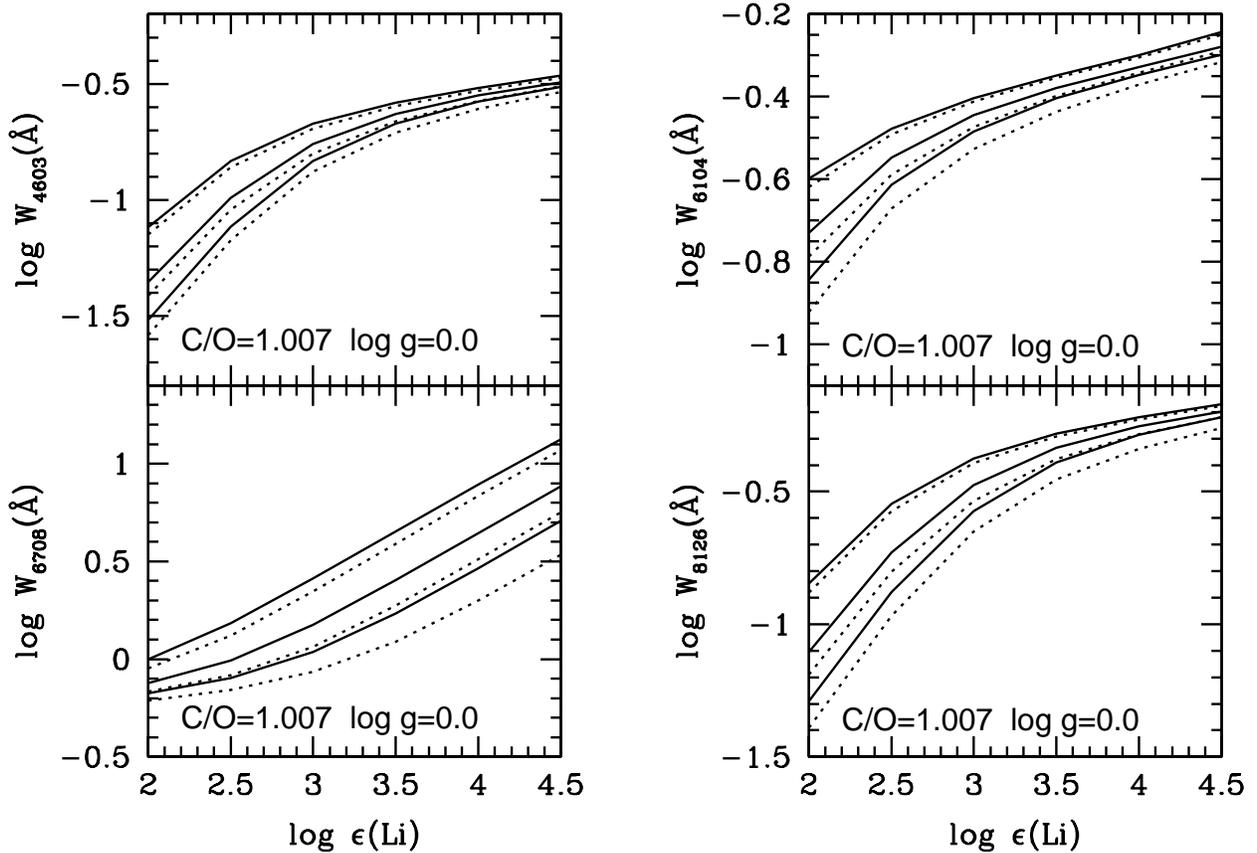}}
\caption{Theoretical LTE (solid) and NLTE (dotted) curves of growth for the four Li I absorptions for
different~\Tef. The pairs of curves (solid-dotted) from top to bottom in each plot are for~\Tef=2500, 2800 
and 3000 K, respectively.}
\end{figure*}

\begin{figure*}
\resizebox{\hsize}{!}{\includegraphics{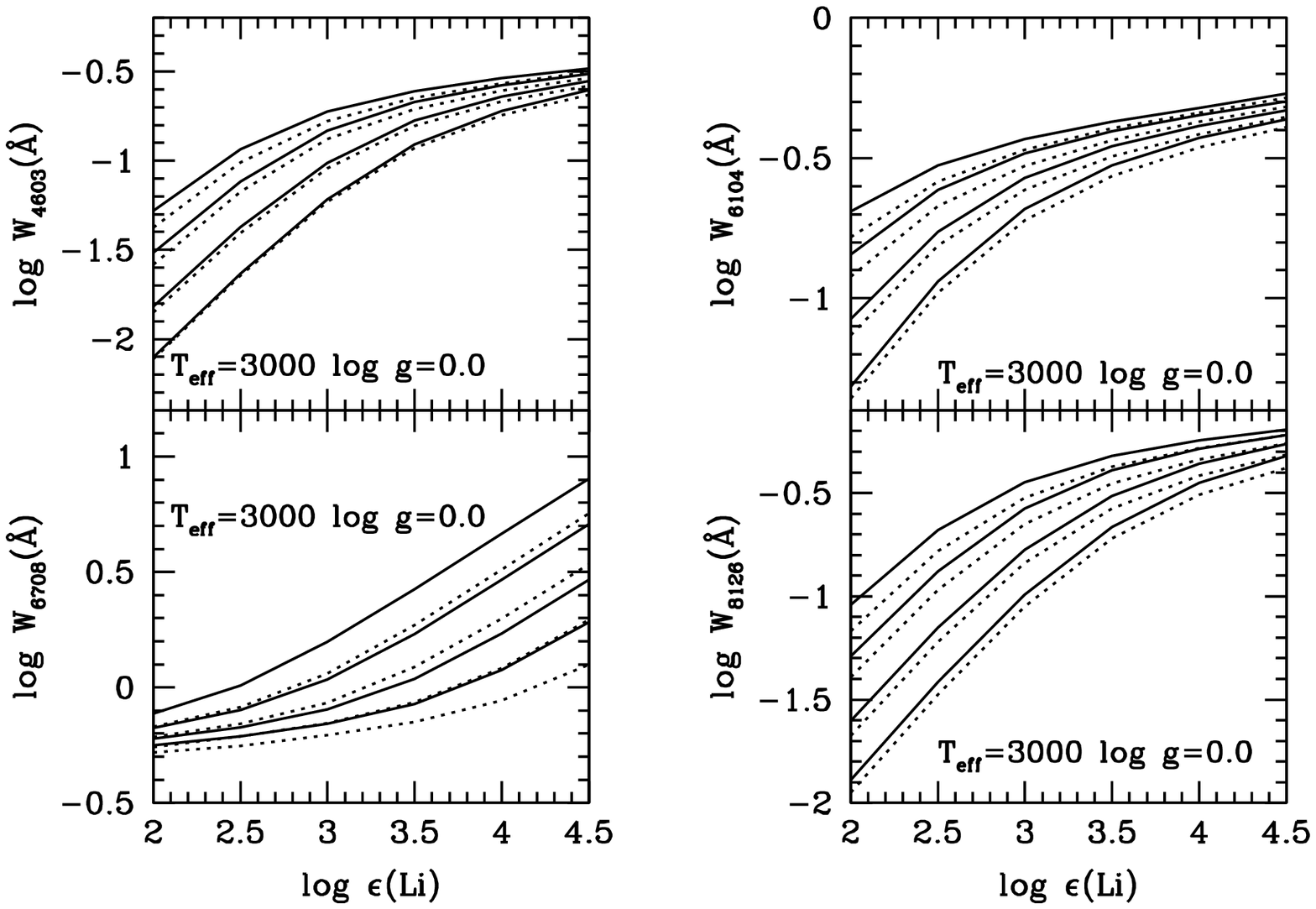}}
\caption{Theoretical LTE (solid) and NLTE (dotted) curves of growth for the four Li I absorptions for
different C/O ratios. The pairs of curves (solid-dotted) from top to bottom in each plot are for C/O=1.007, 1.02, 1.1 
and 1.35, respectively.}
\end{figure*}

Finally, a model atmosphere was interpolated in T$_{\rm{eff}}$ and the C/O ratio for each star from an unpublished 
grid of models for C-stars (Eriksson et al., private communication). These models are constructed under the 
basic hypotheses of hydrostatic equilibrium and plane-parallel approximation. The models include more complete 
opacity data of HCN and C$_2$H$_2$ (in addition to diatomic molecules) with 
the sampling treated by an opacity distribution function (see Eriksson et al. 1984, for details). The final C/O 
ratio of the model atmosphere used in the spectral synthesis 
was obtained by an iterative procedure comparing synthetic and observed spectra until a good fit was 
obtained. Li lines were not included in this procedure. However, the same C/O ratio was not found to give the 
best fit in the four spectral domains for a given star, although differences in the C/O ratio never exceeded 
a few hundredths. For instance, in the $\lambda4603$ and $\lambda6104$ {\AA} regions we systematically 
derived a lower C/O ratio by $\sim 0.03$ hundredths. Table 1 shows the log of the observations and the final 
stellar parameters used in the analysis. The quoted CNO abundances are the mean values of those obtained from  
the best fit for each spectral domain (N abundance plays a minor role).

\subsection{LTE abundances}

Lithium abundances were derived by synthetic spectra {\bf generated by using a modified
version of code written at the Uppsala Astronomical Observatory. The theoretical spectra were} convolved with the
corresponding FWMH to match the instrumental profile in the four spectral domains following the
procedure used in Abia et al. (1993b). 
The carbon isotopic ratios ($^{12}$C/$^{13}$C) derived in Abia \& Isern (1997) from spectral synthesis
to the $^{12}$CN and $^{13}$CN lines in the $\lambda7990-8030$ {\AA} spectral region 
were adopted (see also Abia \& Isern (1996) and de Laverny \& Gustafsson (1998), for details on the analysis 
and the choice of the CN parameters, respectively). Table 1 shows the carbon isotopic ratio adopted 
in the stars. They are in fair agreement with those derived by Ohnaka \& Tsuji (1996) and Lambert et al. 
(1986) in WX Cyg and WZ Cas. The total error due to stellar atmosphere parameter uncertainties,
continuum placement and the fit itself amount to $\pm 0.4-0.5$ dex for Li (LTE) abundances and $\pm 6-13$ for
the $^{12}$C/$^{13}$C ratio (see references above).

Figures 1 and 2 (solid lines) show the behaviour with T$_{\rm{eff}}$ and the C/O ratio of curves of
growth calculated in LTE for the four Li absorptions. As can be clearly seen, subordinate lines
are saturated for high Li abundances (log $\epsilon$(Li)$> 3$). For the resonance line, however, the Li
absorption increases more steeply with increasing Li abundance. For very high Li abundances 
(log $\epsilon$(Li)$>4$), the subordinate lines are not very sensitive to the effective temperature, 
the C/O ratio or gravity (as mentioned before). This is a consequence of their saturated behaviour. The opposite 
occurs with the resonance line. This might lead us to consider the resonance line the better 
tool for abundance determinations in C-stars. However, the resonance line forms in the outermost layers 
(see below) where the atmosphere structure is quite uncertain in C-stars and the microturbulence may 
even become supersonic. We will also see that NLTE effects are much larger for this transition.

Figures 3 to 6 show synthetic fits in LTE to the different spectral domains in WZ Cas. Synthetic spectra 
agree quite well with the observations for the $\lambda6708$ and $\lambda8126$ {\AA} regions but this 
is not the case for the $\lambda4603$ and $\lambda6104$ {\AA} ones. Some of the discrepancy is certainly 
due to the missing features in our line lists as mentioned in 3.1. Moroever, in the $\lambda4603$ and 
$\lambda6104$ {\AA} regions we obtain systematically larger residual fluxes than observed. This, by the way, 
means that the Li abundances derived from the $\lambda4603$ and $\lambda6104$ {\AA} Li lines are systematically lower 
(see Table 3) than those derived from the red Li lines. The same figure was found by de la Reza \& da Silva 
(1995) studying the formation of Li lines in K-giant stars. The reason for this is unknown. 
Bad gf-values for C$_2$ lines, which are numerous in these bluest spectral regions, might partially explain  
this problem. Note that in these two spectral domains most of the C$_2$ lines were taken from the 
data compilation by Kurucz (1998), which is probably not very accurate. Note also that we were not able 
to check their gf values with the solar spectrum as mentioned above. The discrepancy between theoretical and observed 
fluxes is particularly evident in the $\lambda$4603 {\AA} region. We confirmed that most of the strong
absorptions in this spectral range are due to atomic lines while molecular lines contribute most to the
background absorption. However, no reasonable reduction of the metallic abundances in the star can solve 
this discrepancy. Therefore, we are led to believe that an incorrect figure for continuous opacity might 
be the main cause of this. 

The problem of the missing opacity in the blue has been widely discussed in the literature. Recently there 
have been some new results in this sense. Yakovina \& Pavlenko (1998) showed that to fit the head of the 
strong NH band $A^3\Pi-X^3\Sigma^-$~ 
at $\lambda$336 nm in the spectrum of the Sun, one should increase the continuum absorption coefficient 
in the region by a factor $\kappa_\nu\sim 1.7 - 1.9$. Recently, Bell \& Balachandran (1998) increased 
$\kappa_\nu\sim 1.6$ to fit Be II lines at $\lambda$313.0 nm in the solar spectrum. We tried to  
artificially simulate this missing opacity by increasing the H$^-$ opacity by a free factor (namely, increasing the 
electronic pressure by a factor $\sim 3 $). Indeed, in this case theoretical residual fluxes are lower 
(except for the very strong (saturated) lines) and the agreement between observed and theoretical 
spectra improve. We preferred, nevertheless, to be coherent in our analysis in the four spectral domains 
and did not artificially change our set of conventional opacity sources. Thus, we used the same model 
atmosphere for all the spectral ranges and decided, for that reason, not to consider Li abundances derived 
from the $\lambda$4603 {\AA} range. We will have to wait for the next generation of atmosphere 
models for C-stars to study this problem better (Plez et al. 1999). Note that in principle 
sphericity effects should be more pronounced in the blue part of the spectrum because here the sensitivity 
to the temperature structure is higher even for strong spectroscopic features formed in the outer 
atmosphere. 

On the other hand, this discrepancy may be caused (at least partially) by the impact of dust
opacity on the star's spectrum. Indeed, the formation of dust in the atmosphere of AGB stars and
their envelopes has been known for a long time (see references in Wallerstein \& Knapp 1998). Dust particles
can contribute to total opacity via scattering processes, i.e. with cross-sections $\sim 1/\lambda^n$.
In that case the contribution of the dust opacity would increase to the blue. Li-rich K-giants also show 
an red excess, probably caused by a dusty envelope (although of a different nature; de la Reza \& da 
Silva 1995). One may propose other explanations to explain this effect. For instance, the 
procedure used to place the continuum level in the observed spectrum still seems very subjective. To provide
more balanced conclusions regarding this problem we suggest that a more detailed study of these regions should be
carried out in several SLiR stars.

\begin{figure*}
\resizebox{\hsize}{!}{\includegraphics{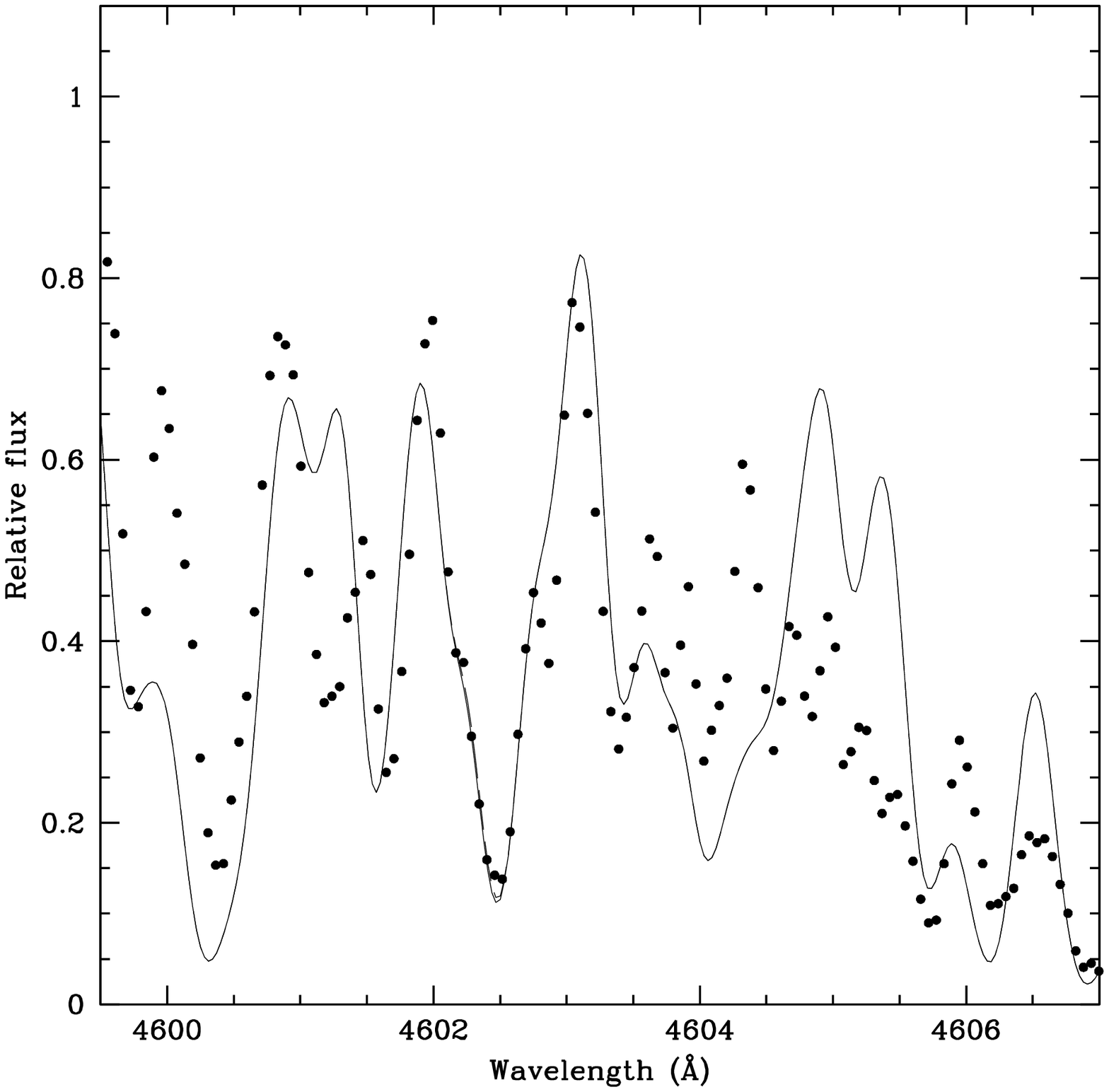}}
\caption{Synthetic spectral fit to the $\lambda4603$ {\AA} spectral domain in WZ Cas for log $\epsilon$(Li)=3.0. Note 
the strong discrepancy between observed (dots) and synthetic spectra in LTE (solid) and NLTE (dashed) (LTE and NLTE
fits to the Li line almost coincide). Most of this
discrepancy is probably due to an incorrect continuous opacity in this spectral range. The Li abundances
derived from this line were not considered.}
\label{}
\end{figure*}

\begin{figure*}
\resizebox{\hsize}{!}{\includegraphics{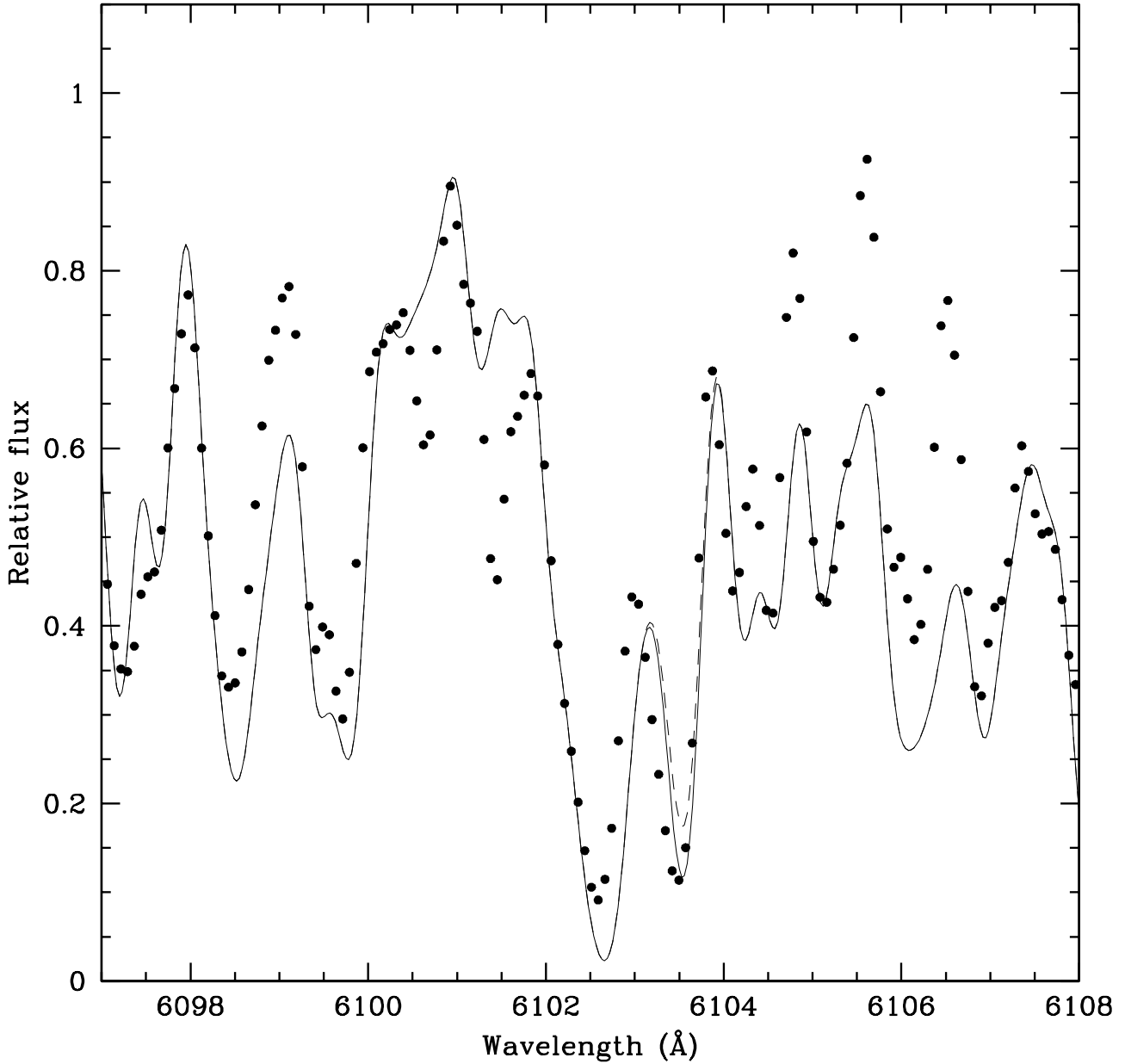}}
\caption{Synthetic spectral fit to the $\lambda6104$ {\AA} spectral domain of WZ Cas for log $\epsilon$(Li)=3.0
LTE (solid line), NLTE (dashed line), observed spectrum (dots).}
\label{}
\end{figure*}

\begin{figure*}
\resizebox{\hsize}{!}{\includegraphics{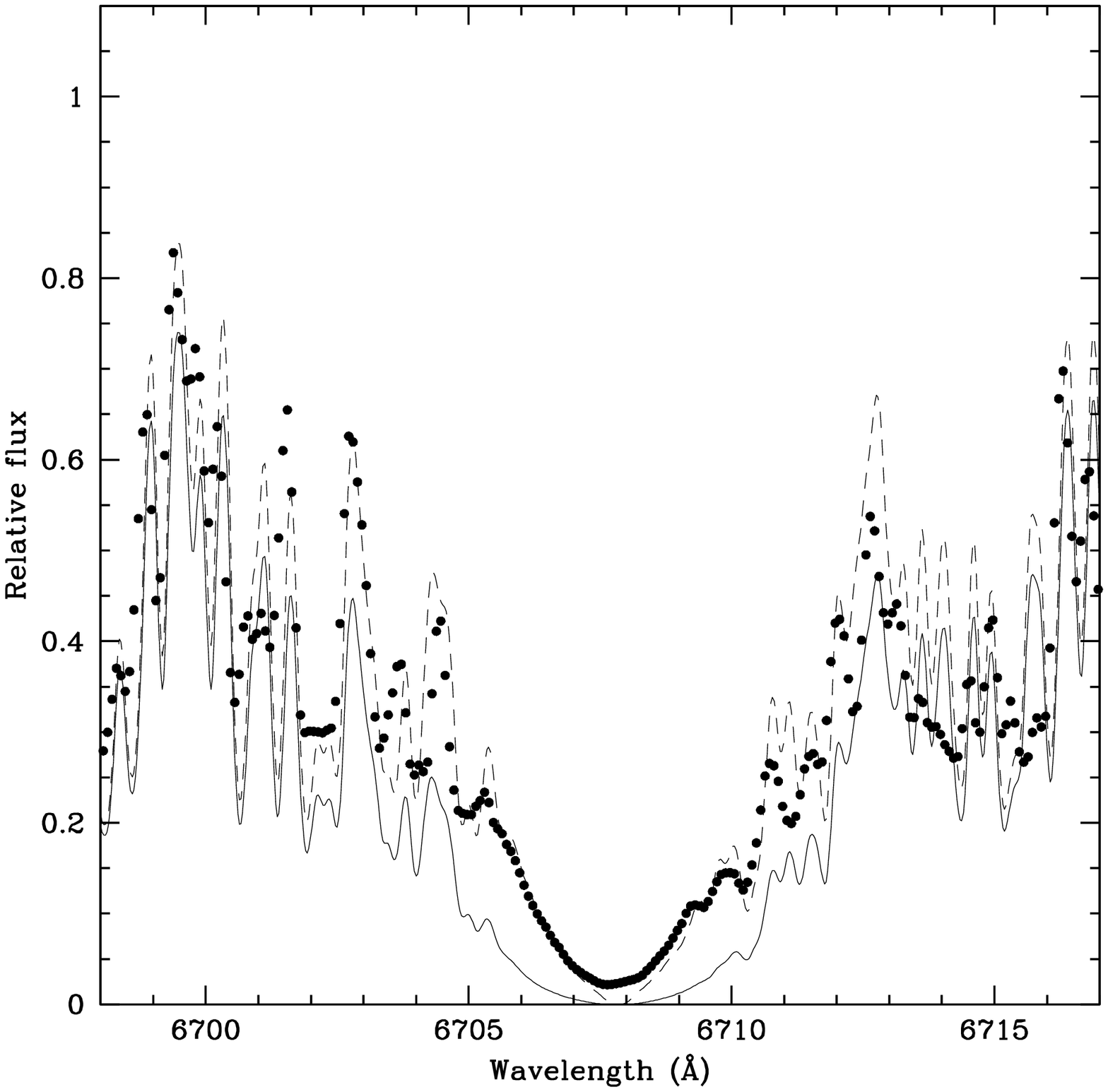}}
\caption{As Figure 4 in the $\lambda6708$ {\AA} spectral domain of WZ Cas for log $\epsilon$(Li)=5.0}
\label{}
\end{figure*}

\begin{figure*}
\resizebox{\hsize}{!}{\includegraphics{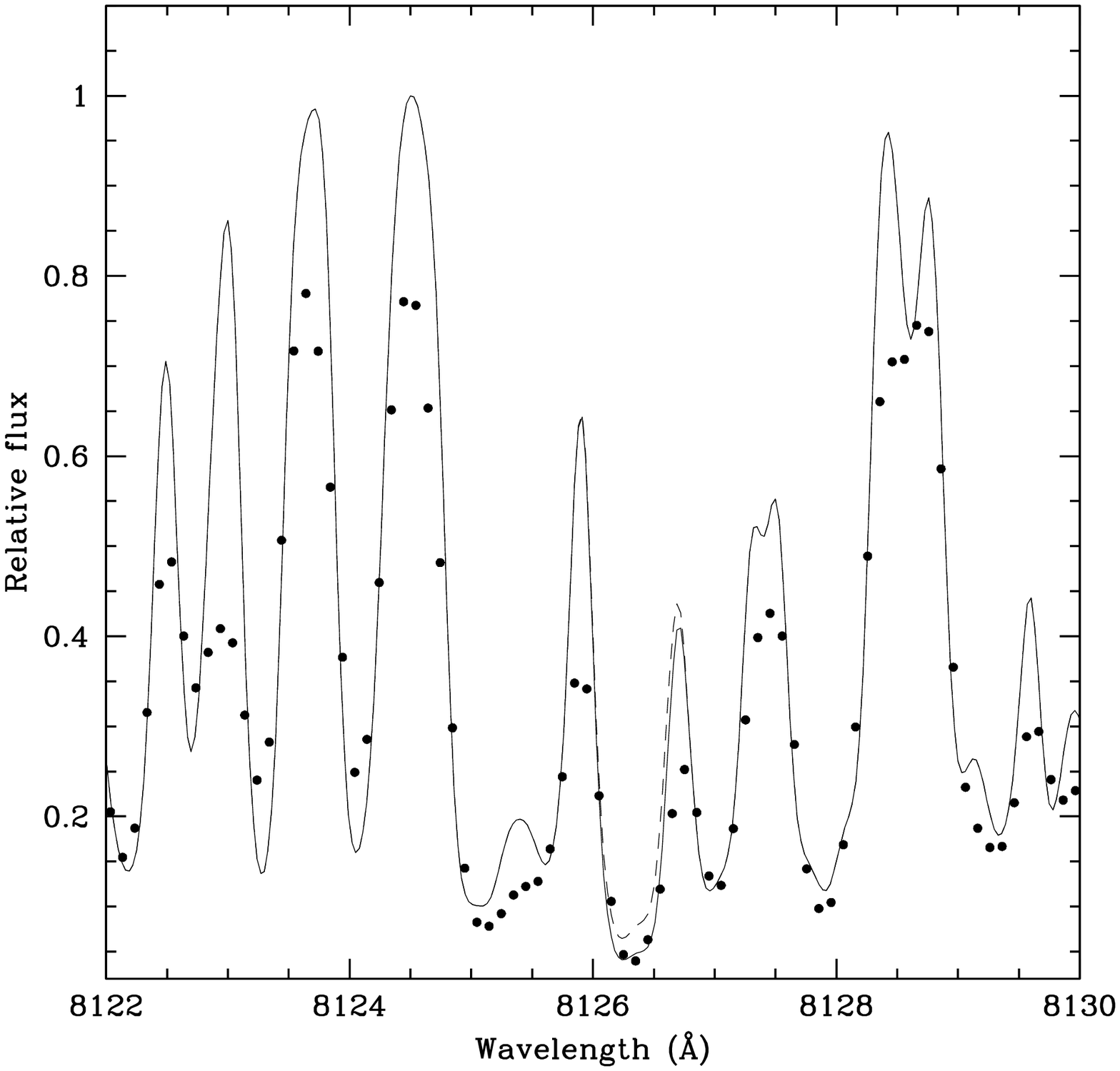}}
\caption{As Figure 4 in the $\lambda8126$ {\AA} spectral domain of WZ Cas for log $\epsilon$(Li)=4.5}
\label{}
\end{figure*}

\subsection{NLTE procedure}

%__________________________________________________ Two column table
\begin{table*}
\caption[]{Data for JOLA opacities used in this work}
{!}{\includegraphics{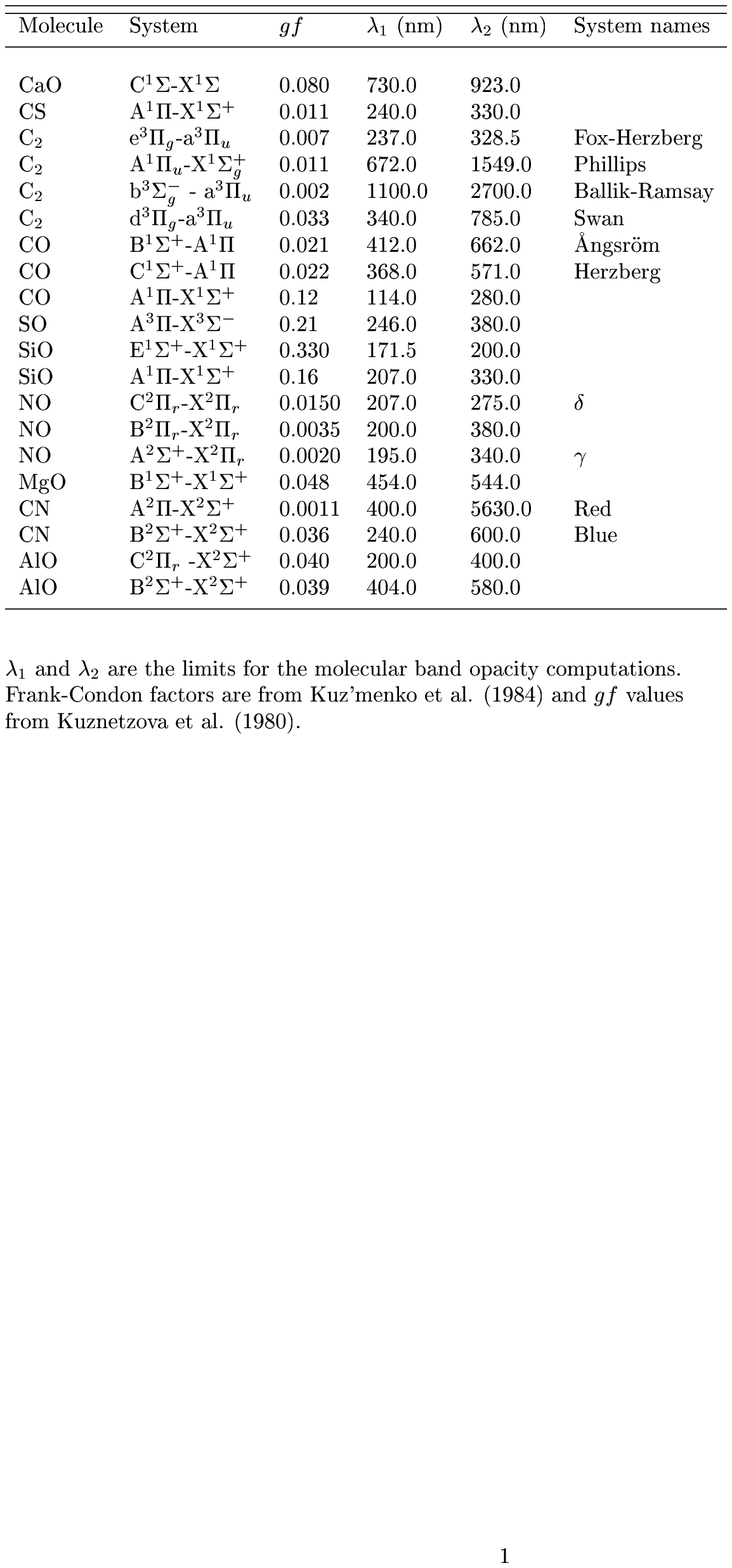}}
\end{table*}

To carry out the NLTE analysis for a 20-level Li atom model, we followed the procedure described in Pavlenko 
(1994) and Pavlenko \& Magazz\'u (1996). A few specific items were taken into account here:
\begin{itemize}
\item ionization-dissociation equilibria were computed for the carbon-rich case, i.e. considering 
C-contained molecules formation: CO, CH, CN, C$_2$, CS, HCN, H$_2$C, H$_3$C, H$_4$C, HC$_2$, 
H$_2$C$_2$ and HC$_3$.     

\item in the frequencies of bound-free transitions of Li I the opacity of diatomic molecules 
was computed in the framework of just-overlapping-line approximation (JOLA, see Table 2).

\item in the frequencies of bound-bound linearized transitions of Li I the atomic absorption 
was included. Furthermore, in the frequencies of the four Li lines we considered the absorption of CN 
and C$_2$ for a given $^{12}$C/$^{13}$C ratio (see Table 1).
\end{itemize}

The work most relevant to the present study was done by de la Reza \& Querci (1978; thereafter RQ). Using a 
divergence flux method these authors carried out an extensive modelling of NLTE effects of lithium lines in 
C-stars. They used a 4 level lithium atom model, atmosphere models by Querci et al. (1974) and Johnson (1974) with
effective temperatures of 3800, 3600 and 3500 K (warmer than the T$\rm{_{eff}}$ values usually derived 
in C-N stars), gravity log g$=1.0$, C/O$=1.3$ and modelled three Li I lines, 
namely the $\lambda6104$, $\lambda6708$ and $\lambda8126$ {\AA} lines. They also tried to model the impact of a
chromosphere on the formation of lithium lines using the approach of {\it radiative temperature} (T$_{\rm{rad}}$) 
on the bound-free transitions of lithium. Several aspects in their NLTE calculation differ from the present study:

\begin{enumerate}
\item {\it The atom model}. We used a 20-level atom model, which allowed us to consider the interlocking of 
lithium lines (transitions) more appropriately. The ionization equilibrium of lithium is formed by the whole 
system of the bound-free transitions. Hovewer, as noted by RQ (see also Pavlenko 1991), transitions 
from/on the second level play the main role. 

\item{\it Bound-free transitions.} The radiation field in bound-free transition computations is very important 
in the modelling of lithium lines without LTE. Indeed, the effectiveness of the overionization of lithium depends 
directly on the mean intensities of the radiation field in the blue part of the spectrum (Pavlenko 1991). 
The approximation $\rm{T_{rad}=T_{e}}$ ({\it electronic temperature}) used by RQ does not produce any overionization 
of lithium. Only in the case of $\rm{T_{rad}>T_{e}}$ does the ionization of lithium increase with respect to the LTE case. 
Note that $\rm{T_{rad}>T_{e}}$ was considered by RQ only when a chromospheric radiation field was included. As noted 
by these authors, this approximation seems to be rather crude for the radiation field.

\item{\it bound-bound transitions}. Lithium lines were treated by RQ as singlets. Basically, radiation transfer in 
the frequencies of several multiplet lines should differ from the case one single (strongest) line. 
Moreover, we include molecular line absorption in the continuous (background) opacity. In principle, 
this should reduce the probability of photon losses from the atmosphere, i.e. it directly affects the processes of 
radiative transfer. On the other hand, the radiative transfer in the continuum differs substantially 
from the case of continuum+lines in an atmosphere with chromosphere, as here the impact of 
the hot chromospheric layers on the photosphere might be weaker. 
\end{enumerate}

\subsection{NLTE curves of growth, synthetic spectra and abundances}
{\bf NLTE computations were performed by using a modified version
of the code WITA2 (Pavlenko et al. 1995) described in Pavlenko (1999)}\footnote{\bf {Comparison between LTE
abundances derived with this code and those obtained with the Uppsala's code showed an excellent agreement
($\sim\pm 0.01$ dex).}}. The dependence of the departure coefficients $b_i = \rm{N^i_{NLTE}/N^i_{\rm LTE}}$, where 
$\rm{N^i_{\rm NLTE}}$ and N$^i_{\rm LTE}$ are the NLTE and LTE populations of the $i$ level, 
upon the depth in a typical C-star atmosphere is shown in Fig. 7. 
The behaviour of the departure coefficients of lower and upper lithium levels differ substantially. The 
populations of the lower levels are reduced by the overionization processes. On the 
contrary, upper levels are overpopulated relative to the LTE case because they are 
more closely linked with the continuum (see also Pavlenko 1994, Carlsson et al. 1994, Pavlenko 
\& Magazzu 1996). Indeed, for log $\epsilon$(Li)$>$ 4.0 the departure coefficients of the first level become 
very low in the outermost layers ($b_1<10^{-4}$ at P$_g< 0.01$), where the NLTE core of 
the Li resonance line forms. 

On the other hand, the depth of formation of lithium resonance and subordinate lines 
is quite different. For large lithium abundances (log $\epsilon$(Li)$> 4.0$) the formation region for the resonance 
line shifts toward the outermost layers, where the temperature drops below 1000 K. In fact, we have to extrapolate the 
model atmosphere structure to log P$_{gas}\approx -5$ to take the line formation consistently into account. 
This produces additional problems in the computation of collisional and radiative rates. Here we are approaching  
the interstellar medium regime, where our approximations for collisional and radiative rates may not be valid. Moroever, 
in this regime the splitting of Li terms into sublevels may also become inportant although, since we are dealing with saturated 
lines, we believe that this effect is not critical. Lithium subordinate lines are formed in deeper layers. For instance, even for 
log $\epsilon$(Li)$> 4.5$ the core of the $\lambda6104$ {\AA} doublet forms in a region with temperature $\sim 2000$ K. 
As a consequence, the NLTE formation regime of this Li line changes dramatically: the interlocking processes 
of radiative transitions (see Magazz\'u et al. 1992; Carlsson et al. 1994) become more important 
than overionization. In this situation, line profiles depend on the behaviour of the source function which is 
affected by the whole set of radiative transitions. In brief, considering the formation of lines in the outermost 
layers we note: i) the opacity here drops sharply due to the low molecular and H$^{-}$ density, ii) the electron density 
is low and its effect on collision rates is small and iii) the electron temperature is also low, so that  
the effectiveness of the overionization $\propto\rm{e^{(-(T_e-T_{rad})/kT)}}$ would be high.

Figures 1 and 2 (dashed lines) also show theoretical NLTE curves of growth for the four Li lines. Departures 
from LTE are very important for the resonance $\lambda6708$ {\AA} line. NLTE corrections for this line can 
amount to 0.6 dex! However, NLTE effects for the subordinate lines are weak ($\leq 0.2$ dex) and even 
decrease for strong Li absorptions (high Li abundances, log $\epsilon$(Li)$\geq 4$). We also see from these 
figures that NLTE effects decrease for decreasing \Tef~and increasing C/O ratio in the model atmosphere. 
When \Tef~drops the opacity of the carbon-contained molecules increases so that the difference between 
T$_{\rm{rad}}$ (of the radiation field in the bound-free frequencies of Li) and T$_e$ decreases. The same 
holds when the C/O ratio increases in the atmosphere. It is interesting to note that LTE/NLTE abundances converge 
for high Li abundances for the subordinate lines whatever the \Tef~and/or the C/O ratio in the atmosphere. 
This gives us a chance to use subordinate lines as a tool for deriving Li abundances in Li-rich AGB stars even 
in the LTE approach, despite their saturated nature. Figures 3 to 6 show synthetic fits in NLTE (only for Li lines; 
i.e. dashed lines) in our stars. The difference between the LTE/NLTE fits to the resonance line is remarkable. For the 
subordinate lines, LTE and NLTE fits differ only slightly.

Table 3 shows the final NLTE Li abundances derived in our stars. NLTE corrections are always positive, ranging from 0.1 
to 0.5 dex. As was shown by Magazz\'u et al. (1992) for the case of G-M stars with solar 
abundances, the significance of NLTE effects depends upon the line strength. Li lines of moderate intensity 
(W$_\lambda >0.2$ {\AA}) form in a region where 
S$_\nu(\tau_{\rm{NLTE}}\approx 1) < \rm{B}_\nu(\tau_{\rm {LTE}}\approx 1)$. 
The NLTE cores of the resonance doublet become stronger than the LTE ones and, as a result, NLTE abundance corrections 
$\Delta$log $\epsilon$(Li)$=$log $\epsilon$(Li)$_{\rm {NLTE}}-$log $\epsilon$(Li)$_{\rm{LTE}}$ are negative. However, in the case of the 
strongest (saturated) Li resonance doublets the difference between LTE and NLTE cores cannot be large (because both approach
to zero) then, NLTE corrections again become positive as in the case of weak lines.
Note that there is a region in the atmosphere where $S_\nu<B_{\nu}$ holds for all except
the $\lambda4603$ {\AA} line (Fig. 8) in the case of intermediate strong lines 
(log $\epsilon$(Li)$\sim 3$). In these regions the interlocking processes of Li radiative transitions
play the main role.

\begin{figure*}
\resizebox{\hsize}{!}{\includegraphics{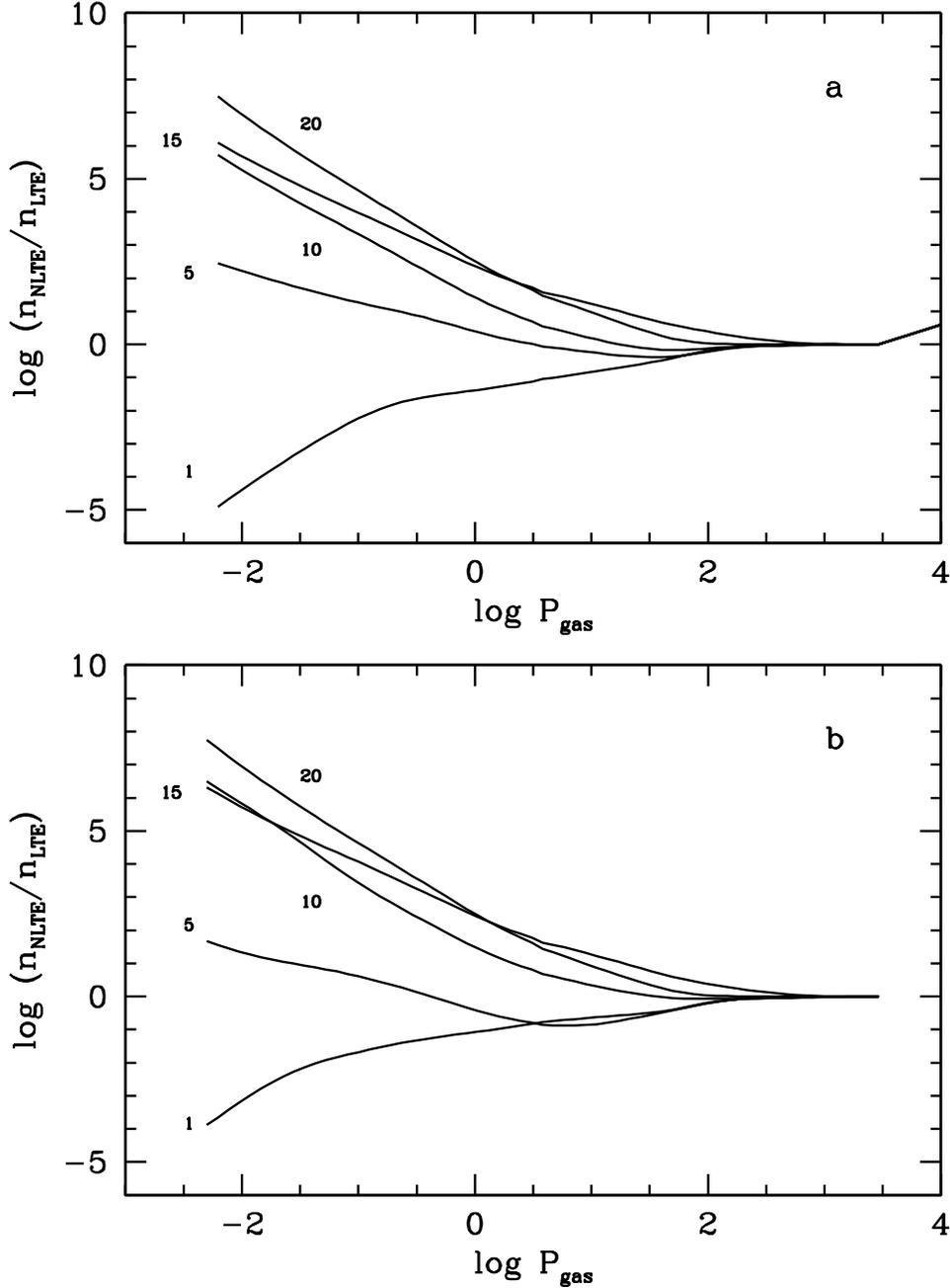}}
\caption{(a) Departure coefficients of five Li levels in the atmosphere of a \Tef/log g/(C/O)$=3000/0.0/1.007$
C-star computed with a lithium abundance of log $\epsilon$(Li)=3.0. (b) The same for log $\epsilon$(Li)=4.5.}
\label{}
\end{figure*}

\begin{figure*}
\resizebox{\hsize}{!}{\includegraphics{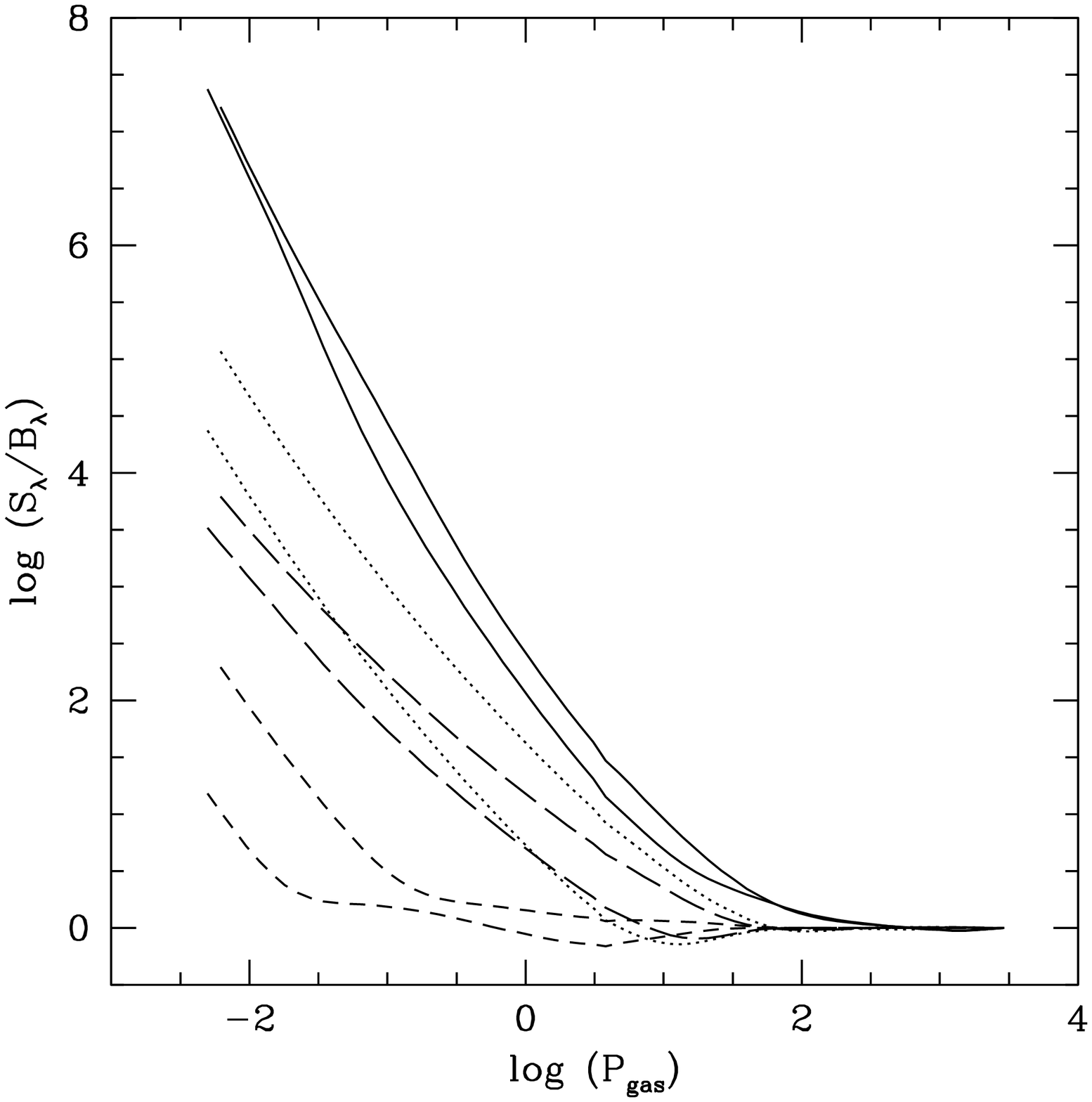}}
\caption{S$_{\nu}$/B$_{\nu}$ ratio for the four Li lines in the atmosphere 3000/0.0/1.007 of a C-star for 
different lithium abundances. Curves correspond to the $\lambda4603$ (solid), $\lambda6104$ (dotted),
$\lambda6708$ (long dashed) and $\lambda8126$ {\AA} (short dashed) Li lines, respectively. For each
pair of curves the upper line is computed with log $\epsilon$(Li)=3.0 and the lower one with log $\epsilon$(Li)=4.5.}
\label{}
\end{figure*}

\section{Discussion and conclusions}

We now compare our LTE/NLTE results with the literature. Abia et al. (1991) derived Li abundances in the
same stars using the $\lambda6708$ Li line. The abundances obtained here are considerably lower. In WZ Cas
and WX Cyg the differences can be ascribed to the different choice of \Tef~and the C/O ratio in the
model atmosphere but also to the more complete molecular line list used in this work, which makes the
background absorption more intense in the $\lambda6708$ {\AA} region. Note that Abia et al.
(1991) did not include the C$_2$ molecule (in any isotopic form) or the $^{13}$CN molecule
in their line list. In IY Hya, moroever, the spectrum now analyzed in the $\lambda6708$ {\AA} 
range shows a much less intense Li line than that from which the Li abundance was derived by these authors:
W$_{6708}\sim 2.5$ {\AA} here against 7.5 {\AA} in Abia et al. (1991). A similar variation in time might be present
in the spectrum of WX Cyg analyzed here (4.5 vs. 6.5 {\AA}). Whether this variation is an effect of a varying \Tef~along the 
pulsational phase or just evidence of the time scale for Li formation/depletion in AGB stars is a 
question which merits further studies. Note that Asplund et al. (1999) found the same figure in a post-AGB star
and they concluded that they were observing the characteristic time scale of the mechanism responsible for the 
lithium production. 

Qualitatively our NLTE results agree with RQ but quantitatively some discrepancies are found.
{\bf For instance, inspection to their Fig. 7 shows a NLTE correction by $\sim 0.7$
dex at log $\epsilon$(Li)$=4$ for the $\lambda 6708$ {\AA} line. However, our computations give a correction of
$\sim 0.4$ dex (see Fig. 2)}. This can be easily explained by differences in the NLTE procedure, model atmospheres etc. 
{\bf Namely}, in this paper we fully computed the opacities in the bound-free and bound-bound 
lithium transitions. This allowed us to establish the 
dependence of NLTE in lithium lines on C/O and the $^{12}$C/$^{13}$C isotopic ratio. Moreover, 
the analysis of RQ was based on the comparison between the LTE and NLTE curves of growths. 
Our LTE/NLTE spectral synthesis approach to determine Li abundances in the atmosphere of 
Li-rich AGB stars seems to be better. Only in this way is the lithium abundance determination possible, since 
the conventional definition of the equivalent width becomes meaningless due to the severe blending of Li
lines. In this sense our results for subordinate Li lines are of great interest. Since we underestimate the 
opacity of the bound-free and bound-bound transitions of lithium we get an upper limit to the NLTE 
effects in these lines, but comparatively, we obtain low NLTE corrections on Li abundances. Using a more 
complete opacity source list, NLTE effects should be even less pronounced. That would comprise a good chance 
to obtain accurate abundance determinations based on the subordinate Li lines in the frame of 
the LTE approach, which is simpler and therefore more useful.
{\bf On the other hand, we can only qualitatively compare our NLTE results with those by Pavlenko \& Magazz\'u (1996) 
in M (O-rich) stars due to the obvious differences in opacities, model atmospheres etc, although
in both cases, strong Li resonance lines are formed in the outermost layers of the atmosphere. Their curves of growth 
(see their Fig. 5) show a similar behaviour than ours (Figs. 1-2): for the $\lambda 6708$ {\AA} line W$_\lambda^{\rm{LTE}}>$
W$_\lambda^{\rm{NLTE}}$ for log $\epsilon$(Li)$>2$; i.e., in the case of strong lines formed high enough in the
atmosphere NLTE effects reduce the equivalent width due to the dominant role of the overionization. 
For the $\lambda 6104$ {\AA} Li line they found W$_\lambda^{\rm{LTE}}<$W$_\lambda^{\rm{NLTE}}$
for log $\epsilon$(Li)$=4.0$ (see their Fig. 6). However, we obtain the contrary. This can be explained
since in our case (C-rich atmospheres) the overionization should increase in the line-forming regions due
to the drop of the density and pressure.}

\begin{table*}
\caption[]{LTE/NLTE lithium abundances in the stars studied}
{!}{\includegraphics{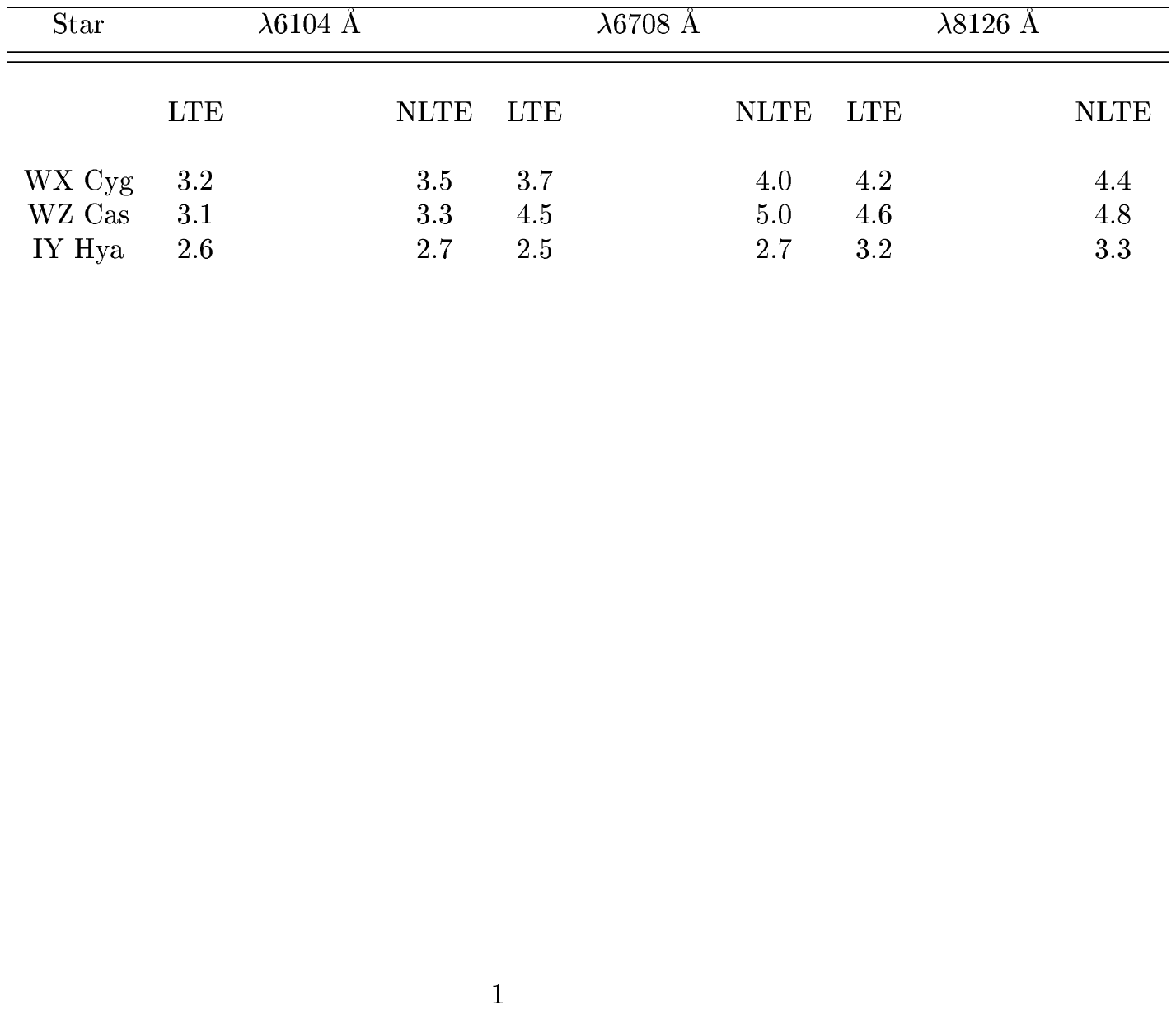}}
\end{table*}

Abia et al. (1993a) computed an empirical Li yield from a homogeneous sample of galactic C-stars. They derived Li
abundances from the $\lambda6708$ {\AA} line in $\sim 220$ C-stars and by considering the mass-loss rate estimates
for the stars in their study, obtained a Li yield into the interstellar medium of $\sim 2\times 10^{-9}$ 
M$_\odot$pc$^{-2}$Gyr$^{-1}$. These authors showed that this yield is extremely dependent on the stellar mass-loss rate assumed 
and on the real Li abundance in the star. Since approximately 90\% of this value is determined by the 
actual yield from the SLiR C-stars, it is straightforward to revise this figure on the basis of the more
accurate Li abundances obtained in this work. Now, considering the NLTE abundances obtained
from the $\lambda8126$ {\AA} Li line as the best estimate of the lithium abundance, we obtain the revised
yield of $\sim 8\times 10^{-10}$ M$_\odot$pc$^{-2}$Gyr$^{-1}$, a factor $\sim 3$ lower than the previous value.
This is mainly because the revised Li abundances in WX Cyg and IY Hya are considerably lower. 
As in Abia et al. (1993a), we can roughly estimate the contribution of C-stars to the galactic Li. Assume that the 
above production rate has been constant during the lifetime of the galaxy ($\sim 13$ Gyr) and that the surface 
density of C-stars has been uniform and also constant during this time within a galactocentric radius of $\sim 15$ kpc 
($\sim 50$ C-stars kpc$^{-2}$, see Claussen et al. 1987). In this case, the total contribution to galactic 
Li abundance by C-stars is M$_{\rm{Li}}\sim 3$ M$_\odot$, i.e. $\sim 10\%$ of the total Li in the galaxy. This contribution 
might be higher if evolutionary effects are taken into account, since the star formation rate was 
certainly higher in the past. Correspondingly, the number of C-stars that eventually became a SLiR C-star at 
a given time was also higher (see Abia et al 1993a for details).

In summary: 

1. The formation of the lithium resonance line in C-stars  
is severely affected by NLTE effects. NLTE Li abundances are higher and can differ from the LTE ones
up to 0.6 dex. The core of this strong Li line is formed
in the outermost layers where we approach the interstellar medium regime. Furthermore,
these layers are affected by different phenomena such as the stellar chromosphere,
inhomogeneities, dusty shells, outflow/infall velocities etc. Our computations also show
that lithium is severely overionized. Therefore, the use of this saturated lithium line for abundance 
determinations in AGB stars seems quite impossible. In that sense, we confirm the results of RQ. 

2. Subordinate lithium lines are formed in the inner parts of the atmosphere. We have shown that 
NLTE effects are rather weak even for strong lines (0.1-0.3 dex). Thus, they may be used for lithium 
abundance determinations in AGB stars even in the framework of the LTE approach. Of the three
subordinate lines, the $\lambda8126$ {\AA} Li I line is probably the best one for Li abundance
determinations because it is less blended, because NLTE effects are weak, because the molecular and atomic 
line list appears to be complete and because continuous opacity seems rather well reproduced 
in this spectral domain.

3. NLTE effects in the lithium lines show a complicated dependence on the input parameters \Tef~and 
the C/O ratio. We found that NLTE effects increase with increasing \Tef~and a
decreasing C/O ratio. For a given \Tef, NLTE effects decrease with increasing Li abundance for the 
subordinate lines but increase for the resonance line. Furthermore, for high Li abundances 
(log $\epsilon$(Li)$>4$) LTE/NLTE abundances from the subordinate lines are not very sensitive to 
\Tef~or the C/O ratio. For the resonance line, however, the opposite happens.

4. The possibility of using the lithium subordinate doublet at $\lambda$4603 {\AA} was considered.
Theoretical spectra, however, show a bad agreement with the observations, probably due to 
a wrong estimate of the continuous opacity in this spectral range and to an incomplete line list in this
region. Furthermore, this lithium line forms a strong blend with iron lines. Due to the very weak
NLTE corrections found for it, the use of this Li doublet would be promising in the case 
of metal deficient stars if using very high resolution spectra. Since we are confident of the atomic and 
molecular line list used beyond $\sim 6000$ {\AA}, we suggest
that the C$_2$ lines of the Swan system observed in the blue part of the spectrum are the largest source of
error. More accurate line lists for these transitions are therefore needed. Such an improvement, together
with better atmosphere modelling, solving the problem of the missing continuous opacity in the blue part
of the spectrum, should improve the agreement between observed and synthetic spectra in the future.

5. We have used observational data of high quality to determine lithium abundances in three 
C-stars: WX Cyg, WZ Cas and IY Hya. We have shown, that they are indeed SLiR stars
with Li abundances in the range log $\epsilon$(Li)$\approx 3$ to 5. However, up to now we have still failed
to accurately determine Li abundances in AGB stars, mainly due to uncertainties in modelling their
atmosphere. This leads to an important uncertainty concerning the estimate of the Li yield by AGB stars. 
These stars might account for up to 30$\%$ of the currently observed Li or merely be a secondary 
source of Li in the galaxy.         

\begin{acknowledgements}      
Patrick de Laverny acknowledges support from the {\it Soci\'et\'e de Secours des Amis des Sciences}. 
Data from the VALD data base at Vienna, Austria, were used for the preparation of this paper. K. Eriksson 
and the stellar atmosphere group of the Uppsala Observatory are thanked for providing the grid of 
model atmospheres. The 4.2 m WHT is operated on the island of La Palma by the RGO in the Spanish 
Observatorio del Roque de los Muchachos of the Instituto de Astrof\'\i sica de Canarias. Based in part on 
observations collected at the German-Spanish Astronomical Center, Calar Alto, Spain. This work
was partially supported by grant PB96-1428.
\end{acknowledgements}

\end{document}